\documentclass[12pt,preprint]{aastex}
\usepackage{graphicx}
\usepackage{caption}
\usepackage{amssymb}
\usepackage{amsmath}
\usepackage{times}
\usepackage{epsfig}

\def\spose#1{\hbox to 0pt{#1\hss}}

\def\multleft#1{\hbox to size{\vbox {\halign {\lft{##}\cr #1}}\hfill}\par}
\def\multright#1{\hbox to size{\vbox {\halign {\rt{##}\cr #1}}\hfill}\par}

\def\boxit#1{\vbox{\hrule\hbox{\vrule\kern3pt\vbox{\kern3pt
          #1 \kern3pt}\kern3pt\vrule}\hrule}}

\makeatletter

\@ifundefined{chapter}{\def\thebibliography#1{
\section*{References}
  \list
  {\relax}{\setlength{\labelsep}{0em}
        \setlength{\itemindent}{-\bibhang}
        \setlength{\itemsep}{\parskip}
        \setlength{\parsep}{0pt}
        \setlength{\leftmargin}{\bibhang}}
    \def\newblock{\hskip .11em plus .33em minus .07em}
    \sloppy\clubpenalty4000\widowpenalty4000
    \sfcode`\.=1000\relax}}%
{\def\thebibliography#1{
  \list
  {\relax}{\setlength{\labelsep}{0em}
        \setlength{\itemindent}{-\bibhang}
        \setlength{\itemsep}{\parskip}
        \setlength{\parsep}{0pt}
        \setlength{\leftmargin}{\bibhang}}
    \def\newblock{\hskip .11em plus .33em minus .07em}
    \sloppy\clubpenalty4000\widowpenalty4000
    \sfcode`\.=1000\relax}}

\begin{document}

\title{The spin-dependence of the Blandford-Znajek effect}

\author{David~Garofalo\altaffilmark{1} }

\altaffiltext{1}{Jet Propulsion Laboratory, California Institute of
Technology, Pasadena CA 91109.  email:
David.A.Garofalo@jpl.nasa.gov}


\begin{abstract}

The interaction of large scale magnetic fields with the event horizon
of rotating black holes (the Blandford-Znajek [1977] mechanism) forms
the basis for some models of the most relativistic jets.  We explore a
scenario in which the central inward ``plunging'' region of the
accretion flow enhances the trapping of large scale poloidal field on
the black hole.  The study is carried out using a fully relativistic
treatment in Kerr spacetime, with the focus being to determine the
spin dependence of the Blandford-Znajek effect.  We find that large
scale magnetic fields are enhanced on the black hole compared to the
inner accretion flow and that the ease with which this occurs for
lower prograde black hole spin, produces a spin dependence in the
Blandford-Znajek effect that has attractive applications to recent
observations.  Among these is the correlation between
inferred accretion rate and nuclear jet power observed by Allen et
al. (2006) in X-ray luminous elliptical galaxies.  If the black hole
rotation in these elliptical galaxies is in the prograde sense
compared with that of the inner accretion disk, we show that both the
absolute value and the uniformity of the implied jet-production
efficiency can be explained by the flux-trapping model.  The basic
scenario that emerges from this study is that a range of intermediate
values of black hole spins could be powering these AGN.  We also
suggest that the jets in the most energetic radio-galaxies may be
powered by accretion onto {\it retrograde} rapidly-rotating black
holes.

\end{abstract}


\section{Introduction}

Astrophysical jets appear to be ubiquitous in accreting systems
surrounded by magnetic fields.  Although the precise mechanism by
which magnetic fields aid in the production of jets is unknown, there
are two basic models that may account for the phenomenon.  The first
involves a magnetocentrifugal or magnetohydrodynamic (MHD) wind
(Blandford \& Payne 1982).  Closely related to the solar wind
phenomenon, an accretion disk threaded by large scale magnetic field
extracts plasma from the disk and expels it along the fieldlines that
extend far away from the source.  Evidence for uncollimated MHD winds
has recently been found in Galactic Black Hole Binaries (GBHBs) by
Miller et al (2006), and such winds may be responsible for the warm
absorber in many active galactic nuclei (AGN; Reynolds 1997; McKernan
et al. 2007).  Despite possessing the appealing feature of being
applicable to outflows from all accreting sources from compact stars
like white dwarfs and neutron stars up to surfaceless objects such as
black holes as well as forming stars, the MHD wind model may fail to
explain the highly relativistic, possibly pair-dominated (Reynolds et
al. 1996; Wardle et al. 1998) jets observed from some AGN.  The second
class of models applies only to black hole accretors, invoking a
direct connection between the magnetic field and the rotating
spacetime of spinning black holes which allows for black hole
spin-energy extraction (Blandford \& Znajek, 1977, henceforth BZ).  In
the BZ mechanism, the power extracted is related to the dimensionless
spin parameter $a$ and the horizon-threading magnetic field $B_H$ via
\begin{equation}
\label{BZ}
L_{BZ} \approx \frac{1}{32}\frac{\Omega_F(\Omega_H-\Omega_F)}{\Omega_H^2}B_H^2r_H^{2}a^{2}c
\end{equation}
where $\Omega_F$ is the angular velocity of observers that measure
zero electric field, $\Omega_H$ is the angular velocity of the black
hole event horizon, $r_H$ is the radius of the event horizon in
Boyer-Lindquist coordinates, and $B_{H}$ is the normal hole-threading
magnetic field, which means that larger hole-threading fields lead to
greater BZ luminosity.  Although this dependence was originally
obtained in the low spin limit, Komissarov (2001) has performed a
numerical study of the BZ solution concluding that the dependence on
spin in equation \ref{BZ} is valid at least up to $a=0.9$. For
completeness, a model that also involves black hole rotational energy
extraction via plasma on negative energy orbits in the ergosphere was
developed by Punsly \& Coroniti (1991); and, a model that involves a
combination of the BZ and BP effects was developed by Meier (1999).

Until the early 1990's, the maximum strength of the hole-threading
magnetic field was determined by considering the strength of the
disk-threading field.  If the horizon-threading field exceeded the
large scale field of the accretion disk, the argument went, the former
would push its way off the hole via magnetic pressure, back into the
disk until the hole-threading field strength was comparable to the
disk field strength.  Two major untested assumptions go into this
scenario.  The first is that the hole-threading field is confined via
Maxwell pressure by the disk-threading field, while the second is that
the disk-threading field grows to a large enough value to confine a
black hole threading field sufficient to drive a powerful jet via the
BZ process.  

In seminal work, Balbus \& Hawley (1991) realized that astrophysical
accretion disks are subject to a powerful MHD instability (previously
discussed by Chandrasekhar [1960] and Velikhov [1959]), and that the
resulting MHD turbulence is likely to be responsible for the angular
momentum transport that facilitates accretion.  However, a consequence
of this realization is that the disk magnetic field must be
characteristically weak (in the sense that gas pressure must exceed
magnetic pressure by an order of magnitude or else the MRI is
suppressed).  This forms the basis for the arguments of Lubow (1994),
Ghosh \& Abramowicz (1997), and Livio et al (1999), pointing to basic
constraints on the strength of the hole-threading field.  Essentially,
these authors argue that if the magnetic field threading the black
hole greatly exceeds the large scale field threading the disk, the
hole-threading field expands back into the accretion disk until an
approximate balance is achieved.  If this were the case, the resulting
weak field threading the black hole would be insufficient to explain
the most powerful AGN jets we see.

In Reynolds, Garofalo \& Begelman (2006; hereafter RGB) we challenged
the argument that Maxwell pressure from the disk solely determines the
field on the hole on the basis that such a conclusion ignores the
dynamics of the accretion flow within the radius of marginal
stability.  Within this radius, circular orbits are no longer stable
and the accretion flow plunges into the black hole.  Hereafter we
refer to the region within the radius of marginal stability as the
``plunge region''.  In RGB, we argue that the inertial forces within
the plunge region prevent magnetic field that is threading the black
hole from expanding back into the disk.  Accretion of magnetic field
can result in a strong flux-bundle threading the black hole, confined
in the disk plane by the plunge region.  This flux bundle does,
however, expand at high latitudes due to Maxwell pressure.
Steady-state is obtained when field lines threading the inner disk are
bent by the high latitude regions of the flux bundle such that
tension-induced outwards diffusion of the field line through the
accretion disk balances inwards advection of the field line.

In this paper, we adopt the formalism of RGB and extend it to the
relativistic regime in order to determine the spin dependence of
magnetic flux accumulation on the black hole.  We also show that the
$a=0$ limit produces flux accumulation values close to those obtained
in the non-relativistic study, establishing the validity of the RGB
results for slowly spinning black holes.  Our central result is that
the ability of the plunge region to enhance the black hole threading
field decreases as the spin of the black hole increases.  Despite this
decrease in the ability of the plunge region to produce a black
hole-threading field that is enhanced with respect to the inner
accretion flow, the hole-threading field is always greater than the
field strength in the inner accretion flow.  This means that the BZ
power in our model is always greater than in ``classical'' BZ where
this enhancement is not included.  In Section \ref{relativity}, we
describe the formalism of the relativistic extension of the
flux-trapping model.  We discuss the covariant nature of the magnetic
flux function and the equations it satisfies.  In Section
\ref{results} we present our results and show that the hole-threading
flux decreases with increase in spin for lower prograde spins.  This
appears to be largely a geometrical effect connected to the radial
position of the marginally stable orbit and the horizon as a function
of spin.  In section 4, we discuss the implications of our model in
view of the recently discovered correlation between jet power and
accretion rate found by Allen et al. (2006).  Our model also predicts
that accretion onto rapidly rotating {\it retrograde} black holes will
produce extremely powerful jets, and we suggest this as a mechanism
for powering the most luminous radio-loud AGN.  Section
\ref{conclusion} presents our conclusions.

\section{Relativistic Generalization of the Flux-Trapping Model}
\label{relativity}

Our goal is to construct a relativistic version of the model described
by RGB.  We consider a geometrically-thin accretion disk around a Kerr
black hole threaded by a large scale magnetic field.  On
small scales, the magnetic field lines are frozen into the highly
conductive plasma of the accretion disk.  However, if we coarse-grain
our view to larger scales, we expect that the large scale magnetic
field lines can undergo turbulent diffusion through the disk plasma.
Heyvaerts et al. (1996) have shown that the effective magnetic Prandtl
number (i.e., the ratio of the effective turbulent viscosity to the
effective turbulent magnetic diffusivity) is expected to be within an
order of magnitude of unity.  Hence a field line threading the disk
will be dragged inwards by accretion, but radial magnetic pressure
gradients and magnetic tension (associated with field line curvature
as it threads the disk) will lead to competitive field line diffusion.
We assume the region above and below the disk has a very low
plasma density, and hence for the magnetic field in this region to
have a force-free configuration.  As argued by RGB, field
lines threading through the plunge region of the accretion flow will
be dragged very rapidly onto the black hole, leaving the plunge region
devoid of poloidal magnetic flux.

Below we describe the idealizations and construction of our model
system. 
\begin{enumerate}
\item Our accretion disk is described by a Novikov \& Thorne (1974)
disk truncated at the marginally stable orbit, inwards of which is the
plunging region.  We assume that the large scale magnetic field does
not perturb the structure of the turbulent accretion flow.

\item In the magnetosphere (the region outside of the black hole and
accretion disk) we assume that the plasma density is negligible and
hence that the magnetic field is force free and we impose a current
density distribution that is compatible with the horizon regularity
condition (see below).

\item As discussed above, we assume that no poloidal magnetic flux threads the
plunge region of the accretion disk.  Any magnetic flux that is
advected inwards across the radius of marginal stability is
immediately added to the flux bundle threading the black hole.  This
is the distinguishing fundamental assumption of our model. 

\item The boundary condition on the horizon requires imposing finite
  electric and magnetic fields as measured by freely-falling observers
  crossing the horizon.  Znajek (1978) worked out the appropriate
  boundary, or regularity, condition on the horizon under this assumption for a
  force-free magnetosphere.  This condition is imposed on the horizon
  in our model.  

\item Far away from the black hole and at poloidal angles above the
  accretion disk, we assume the large-scale field is uniform and
  impose an appropriate outer boundary condition that captures this
  assumption.  Far away from the black hole but in the plane of the
  accretion disk we impose a ``dead zone'' as in RGB.  The only
  difference between this region and the active disk region is that
  the radial inflow velocity is set to zero.  In the dead zone,
  therefore, magnetic field lines are not dragged toward the black
  hole.  As pointed out in greater detail in RGB, this dead zone
  dramatically reduces the sensitivity of the system to the treatment
  of the outer boundary.  The physical nature of the dead zone which
  is addressed in RGB, allows us to smoothly truncate the accretion
  disk. The outer boundary condition applied here amounts to bounding
  the system with a perfectly conducting sphere.  This differs from
  RGB who, due to the analytic techniques possible in the
  non-relativistic problem, could bound the system in the plane by a
  large conducting annulus, and otherwise just apply boundary
  conditions at infinity out of the plane.

\end{enumerate}

\begin{figure*}[h!]
\centerline{\includegraphics[angle=0,scale=0.6]{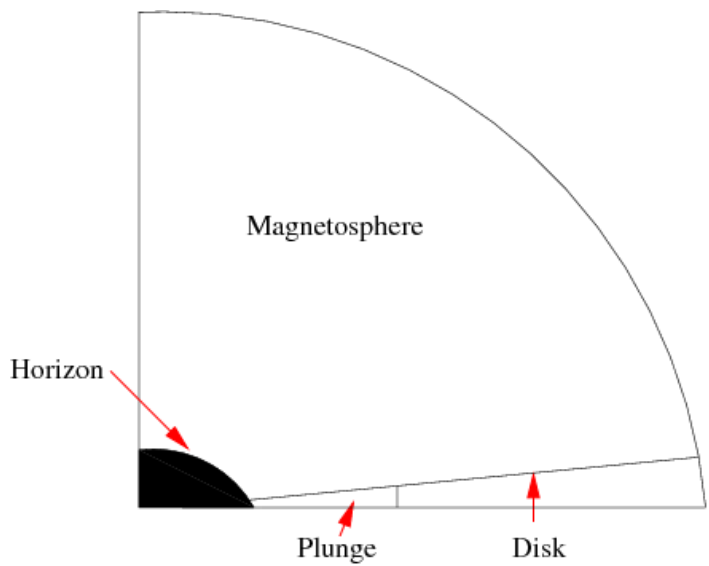}}
\caption{Boundaries for the axisymmetric numerical
solutions}\label{boundaries}
\end{figure*}

\subsection{Basic equations of our model}

We assume that the underlying spacetime is that of an isolated
rotating black hole, i.e., that the gravitational field of the
rotating black hole completely dominates over that of the accretion
disk.  Throughout this paper, we will work in Boyer-Lindquist
coordinates in which the Kerr metric takes the standard form,
\begin{eqnarray}
  dS^{2}&=&-\left(1-\frac{2Mr}{\rho^{2}}\right)
  dt^{2}-\frac{4Mar\sin^{2}\theta}{\rho^{2}}dt\,d\phi\\
  &+&\frac{\Sigma}{\rho^{2}}\sin^{2}\theta\, \nonumber
  d\phi^{2}+\frac{\rho^{2}}{\Delta}dr^{2}+\rho^{2}d\theta^{2},
\end{eqnarray}
where 
\begin{equation}
\rho^{2} = r^{2} + a^{2}\cos^{2}\theta,
\end{equation}
\begin{equation}
\Delta = r^{2}-2Mr+a^{2},
\end{equation}
and
\begin{equation}
\Sigma = (r^{2}+a^{2})^{2}-a^{2}\Delta\sin^{2}\theta.
\end{equation}

The basic equation describing the evolution of the (large-scale)
magnetic field within the acccretion disk is obtained by following the
relativistic analogue of the non-relativistic treatment of RGB.  One
difference between our approach and that of RGB is that we only seek
the stationary (time-independent) solution; this results in a
significant simplification in the relativistic equations that we must
deal with.  The equation describing the field evolution within the
disk is obtained by combining Maxwell's equation
\begin{equation}
\triangledown_{b}F^{ab} = \mu J^{a},
\end{equation}
with a simplified Ohm's law
\begin{equation}
\label{Ohm}
J^{a} = \sigma F^{ab}u_{b} - u^{a}J_{b}u^{b},
\end{equation}
where $F^{ab}$ is the standard Faraday tensor, $\mu$ is the
permeability of the plasma, $J^a$ the 4-current, $u^a$ the 4-velocity
of the accretion disk flow, and $\sigma$ the effective conductivity
of the turbulent plasma.  This gives
\begin{equation}
\triangledown_{b}F^{ab} = \frac {1}{\eta} F^{ab}u_{b}- \mu
u^{a}u_{b}J^{b},
\label{disk}
\end{equation}
where $\eta = 1/\mu\sigma$ is the effective magnetic diffusivity.  To
reiterate, the microscopic magnetic diffusivity of the plasma is
expected to be essentially zero.  However, in a coarse-grained view,
magnetic field lines will diffuse through the turbulent plasma due to
small scale reconnection events, and it is this process that is
parametrized through the effective magnetic diffusivity.
The second term on the right-hand side of equation \ref{disk}, which
we include for completeness, is zero by the MHD assumption that proper
electric charge vanishes.  Therefore, the disk is governed by
\begin{equation}
\triangledown_{b}F^{ab} = \frac {1}{\eta} F^{ab}u_{b}.
\label{disk1}
\end{equation}
The equations are cast in terms of the vector potential, which is
related to the Faraday tensor via
\begin{equation}
 F_{ab}=A_{b,a}-A_{a,b},
\label{vec_pot}
\end{equation}
and, in particular, in terms of the component $A_{\phi}$ in the
coordinate basis of the Boyer-Lindquist coordinates.   

Ultimately, to examine BZ powers, we need to derive the magnetic flux
threading a hoop placed at a given radius $r$.  The magnetic flux
function is related to the vector potential via Stokes' Theorem
applied to the Faraday tensor
\begin{equation}
\psi \equiv \int_{S}F=\int_{S}dA=\int_{\partial{S}}A=2 \pi A_{\phi}, 
\end{equation}
where $S$ is a space-like surface with boundary $\partial S$
consisting of a ring defined by $r={\rm constant}$, $\theta={\rm
constant}$, and $t={\rm constant}$.  Because we work with the vector
potential, $A_{b}$, we comment briefly on the choice of gauge.  Since
$A_{b}$ is specified up to the gradient of a scalar function $\Gamma$,
\begin{equation}
A_{b}^{'} = A_{b} + \triangledown _{b} \Gamma,
\end{equation}  
the assumption of time-independence and axisymmetry gives us
\begin{equation}
A^{'}_{t} = A_{t}
\end{equation}
and
\begin{equation}
A_{\phi}^{'} = A_{\phi}.
\end{equation}
Thus, we need not specify the gauge uniquely beyond the statement of t
and $\phi$ independence.  Writing the $\phi$ component of
eqn. \ref{disk1} in terms of the vector potential, and applying
time-independence and axisymmetry yields,
\begin{eqnarray}
\label{disk2}
\frac{\partial}{\partial r}\left[g^{11}\left(g^{30}\frac{\partial
A_{t}}{\partial r}+g^{33}\frac{\partial A_{\phi}}{\partial
r}\right)\right] + \frac{\partial}{\partial
\theta}\left[g^{22}\left(g^{30}\frac{\partial A_{t}}{\partial
\theta}+g^{33}\frac{\partial A_{\phi}}{\partial \theta}\right)\right]
+\\\nonumber \frac{1}{2}\left(g^{00}\frac{\partial g_{00}}{\partial
r}+2g^{30}\frac{\partial g_{30}}{\partial r}+g^{33}\frac{\partial
g_{33}}{\partial r}+g^{11}\frac{\partial g_{11}}{\partial
r}+g^{22}\frac{\partial g_{22}}{\partial
r}\right)g^{11}\left(g^{30}\frac{\partial A_{t}}{\partial
r}+g^{33}\frac{\partial A_{\phi}}{\partial r}\right) +\\\nonumber
\frac{1}{2}\left(g^{00}\frac{\partial g_{00}}{\partial
\theta}+2g^{30}\frac{\partial g_{30}}{\partial
\theta}+g^{33}\frac{\partial g_{33}}{\partial
\theta}+g^{11}\frac{\partial g_{11}}{\partial
\theta}+g^{22}\frac{\partial g_{22}}{\partial
\theta}\right)g^{22}\left(g^{30}\frac{\partial A_{t}}{\partial
\theta}+g^{33}\frac{\partial A_{\phi}}{\partial \theta}\right)
\\\nonumber = \frac{1}{\eta}u_{r}g^{11}\left(g^{30}\frac{\partial
A_{t}}{\partial r}+g^{33}\frac{\partial A_{\phi}}{\partial r}\right),
\end{eqnarray}
where the $g_{\alpha \beta}$ and $g^{\alpha \beta}$ are the lower and
upper metric terms in the Boyer-Lindquist coordinates, and are
evaluated at the disk surface [$\theta=\pi/2-\tan^{-1}(h/r)$ where
$h/r$ is the fractional thickness of the disk].      This is our final
equation describing the magnetic flux threading the accretion disk.  

As in the non-relativistic case (RGB), we need to match the field in
the disk onto the magnetospheric field in order to fully specify the
solution.  As described previously, our assumptions for the
magnetosphere lead to the force-free condition
\begin{equation}
F^{ab}J_{b} = 0
\label{force-free}
\end{equation}
and
\begin{equation}
\triangledown_{b}F^{ab} =  \mu J^{a}.
\label{magnetosphere1}
\end{equation}
In the magnetosphere we also impose the ideal MHD condition
\begin{equation}
F^{ab}u_{b} = 0,
\label{ideal}
\end{equation}
where $u^b$ is the 4-velocity of the (tenuous) plasma in the
magnetosphere and is determined by the condition that (in
steady-state) field lines rigidly rotate. 


Our basic philosophy is to solve eqn.~\ref{magnetosphere1}
supplemented by the constraints given by equations \ref{force-free}
and \ref{ideal} for the magnetic field structure in the magnetosphere
using eqn.~\ref{disk2} as a boundary condition to be applied on the
disk surface.  Additional boundary conditions are required.  The
magnetic flux is fixed to be zero on the black hole spin axis (i.e.,
the field is assumed to be finite on the axis).  We bound the region
under consideration by an outer spherical boundary at large $r$, and
assume that the flux threading that boundary corresponds to a uniform
field with strength $B_0$, i.e., we set $ \psi = r^{2}\sin ^{2}\theta
B_{0}$.  The regularity condition on the horizon is determined by the
Znajek condition (Znajek 1978; Macdonald 1984) with explicit form,
\begin{equation}
\label{Znajek}
\psi = \frac{2\psi_{0}}{1 +
  \left[\sin^{2}\theta/(1-\cos\theta)^{2}\right]\exp[-2a^{2}\cos\theta/(r_{+}^{2} + a^{2})]},
\end{equation}
where $r_{+}$ is the radial coordinate of the horizon and $\psi_{0}$
is the magnitude of the flux threading the horizon which is determined
by the numerical solution.  In essence, the regularity condition
imposed on the horizon amounts to the $a$ and $\theta$ dependence of
the above function only.  Finally, the fundamental assumption of our
model is that no poloidal magnetic flux threads the plunging region of
the disk ($r_{\rm evt}<r<r_{\rm ms}$; $\theta=\pi/2-\tan^{-1}(h/r)$).
The prescription for the 4-velocity comes from the assumption that
flux tubes rigidly rotate.  Flux tubes that intersect the black hole
all rigidly rotate but at values that depend on the polar angle
whereas flux tubes that intersect the disk rotate rigidly at a
Keplerian frequency determined at the disk surface except for the
inner disk where the plunge region approaches.  The prescription of
field line rotation is such that at large polar angles, the field
lines threading the horizon rotate at values that approach those of
the field lines that intersect the inner accretion flow.  These
features exist to avoid the presence of discontinuities in the current
prescription.  The current prescription, on the other hand, is fixed
on the polar axis and needs to be determined on the disk surface as
the flux function does via a relaxation approach.  In general, on the
horizon and outer boundary, a regularity condition needs to be imposed
(Macdonald \& Thorne, 1982; Uzdensky, 2005).  However, the
prescription for the rigid rotation of flux tubes is chosen to avoid
having to deal with an outer light cylinder, thereby allowing one to
provide a simple boundary condition for the current as done for the
flux function itself.  This prescription simplifies calculations.  On
the horizon, instead, a regularity condition must be enforced which
means that the prescription in the magnetosphere must be relaxed along
with the regularity condition on the horizon until the two are
compatible.

\subsection{Solution method}

We adopt a relaxation method approach to solve for the
time-independent magnetic flux configuration around a Kerr black hole.
Through this relaxation process, we derive the steady-state solution
to equation \ref{magnetosphere1} subject to constraints
\ref{force-free} and \ref{ideal} given the boundary conditions
discussed above.  At the start of the numerical solution, we thread
the accretion disk with uniform magnetic field everywhere which is not
a steady-state solution to equation \ref{disk2}.  We then jointly
relax the magnetic configuration both in the magnetosphere and on the
disk surface until a solution to equations \ref{magnetosphere1} and
\ref{disk2} is obtained.  As the solution evolves, the magnetic flux
at the disk inner edge changes.  As previously mentioned, a
consequence of our boundary condition on the plunge region is that any
magnetic flux advected across the radius of marginal stability is
immediately added to the flux bundle threading the black hole.  As the
disk supplies flux to the horizon via the plunge region, Maxwell
pressure will lead to a high-latitude expansion of the hole-threading
flux bundle, changing the field geometry in the magnetosphere away
from the uniform initial state.  As this happens, the diffusion terms
in equation \ref{disk2} increase.

Flux accumulation occurs even once the hole-threading field is
significantly greater than the disk-threading field because the plunge
region is shielding the disk-field from the magnetic pressure
associated with the hole-threading field.  However, the system does
settle into steady-state when the disk-threading field is bent by the
expanded black hole flux-bundle such that outward field line diffusion
balances inward advection.  The physics is similar to that described
in RGB with the exception of the presence of finite poloidal currents
in the magnetosphere as well as relativistic vs. Newtonian spacetime.

In our canonical numerical solution, space is divided into a
$(r,\theta)$-grid, with 72 zones in $r$ and 51 zones in $\theta$.  The
radial coordinate runs from the horizon to an outer boundary at
$r=53$, and is spaced in a geometric progression such as to give a
factor of almost 2 difference in the zone spacing at the inner and outer
boundary.  The $\theta$ coordinate runs from the axis ($\theta=0$) to
the disk surface, and is uniformly spaced in $\cos\theta$.

\subsection{Newtonian vs. relativistic treatments of non-rotating black holes}
\begin{figure*}[h!]
\centerline{\includegraphics[angle=-0,scale=0.6]{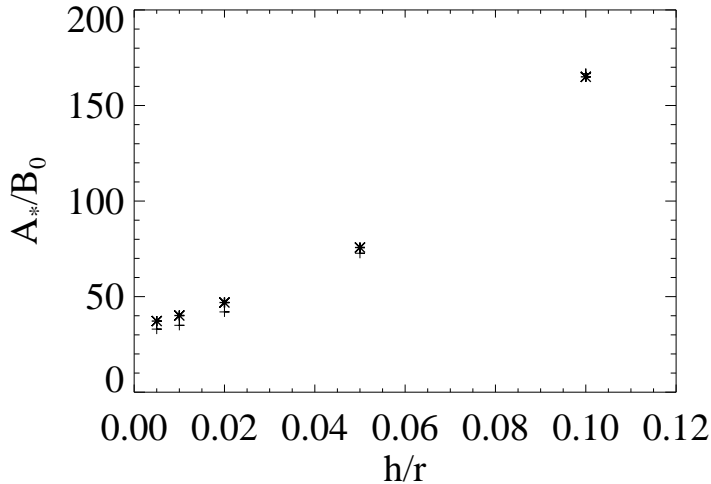}}
\caption{Equilibrium value of $A_{*}$/$B_{0}$ as a function of h/r for
  $Pr_{m}$=20.0 for Newtonian (asterisk) and Schwarzschild disk
  (cross).  Notice how the Newtonian values are larger and by a
  greater amount for smaller disk thickness.  As the thickness
  increases the difference between Newtonian and Schwarzschild
  decreases until for largest thickness the Schwarzschild value
  overtakes the Newtonian.  Nevertheless, the differences are always
  of a few percent, establishing the fact that the Newtonian treatment
  is sufficiently accurate for slowly rotating black
  holes.}\label{hoverr}
\end{figure*}

In this section we show that the Newtonian analysis of RGB accurately
describes the physics of the flux-dragging model in the slowly
rotating black hole case.  We choose to compare the flux-trapping that
results for $a=0$ against a Newtonian treatment, with magnetic Prandtl
number fixed at $Pr_{m}=20$ and varying disk thickness.  Given our
different treatment of the outer boundary condition, it would be
inappropriate to compare our Schwarzschild results directly with the
results of RGB, however.  Instead, we use our code to derive both the
Schwarzschild results and the Newtonian limit ($c$$\to \infty$).  The
results are displayed in Fig. \ref{hoverr}, where, to be consistent
with the notation of RGB, we plot the ratio $\frac{A_{*}}{B_{0}}$
where $A_{*}$ = $A_\phi$.  The difference between the Newtonian and
relativistic cases for all disk thicknesses is a few percent.  This
demonstrates that the neglect of the relativistic terms for slowly
rotating black holes is justified, as hypothesized by RGB.

\subsection{Resolution and convergence study}
\begin{figure*}[h!]
\centerline{\includegraphics[angle=-0,scale=0.6]{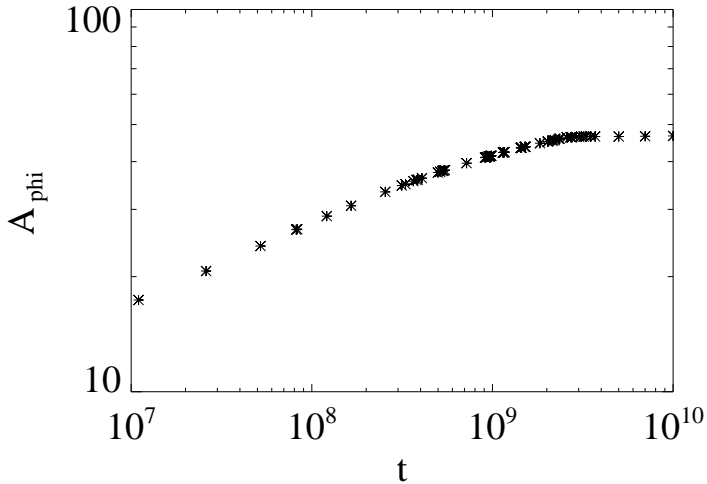}}
\caption[Resolution figure.]{Value of $A_{\phi}$ as a function of the
relaxation parameter $t$ showing the typical behavior of the runs.
This one is for $a=0.2$.}\label{convergence}
\end{figure*}

It is necessary to confirm that our results do not rely on our
particular choice of numerical resolution.  We chose to examine the
effects of resolution on a representative model with $a=0.4$,
$h/r=0.1$, and $Pr_{m}=20$, increasing in resolution from our
canonical 72 by 51 case to 144 by 102.  The results can be seen in
Table 1, where we compare the equilibrium hole-threading flux of each
case with that obtained at our canonical resolution.  It can be seen
that the higher resolution run agrees very well with a hole-threading
flux that is only $1.5\%$ lower despite quadrupling the number of
computational cells.
\begin{table}
\begin{center}
\begin{tabular}{||c|c|c|c|c||}
\hline
spin & $r\times \theta$  grid     &     $\psi_{norm}$     &   \% difference \\ \hline
0.4  &  72  by 51   &     1           &   0             \\ \hline
0.4  & 100  by 51   &     1.004       &   0.4           \\ \hline
0.4  & 100  by 80   &     0.985       &   1.5            \\ \hline
0.4  & 144  by 102  &     1.01        &   1.1            \\ \hline
\end{tabular}
\end{center}
\caption{Resolution study for flux obtained at $a=0.4$.  }
\end{table}
We then show the convergence to the relaxation solution.  In
Figure \ref{convergence} we show the dependence of the flux function
on the relaxation parameter $t$ for the specific value of $a=0.2$.
This behavior is typical for all spin runs.

\section{Results: spin dependence of flux trapping} 
\label{results}
\begin{figure*}[h!]
\centerline{\includegraphics[angle=-0,scale=0.8]{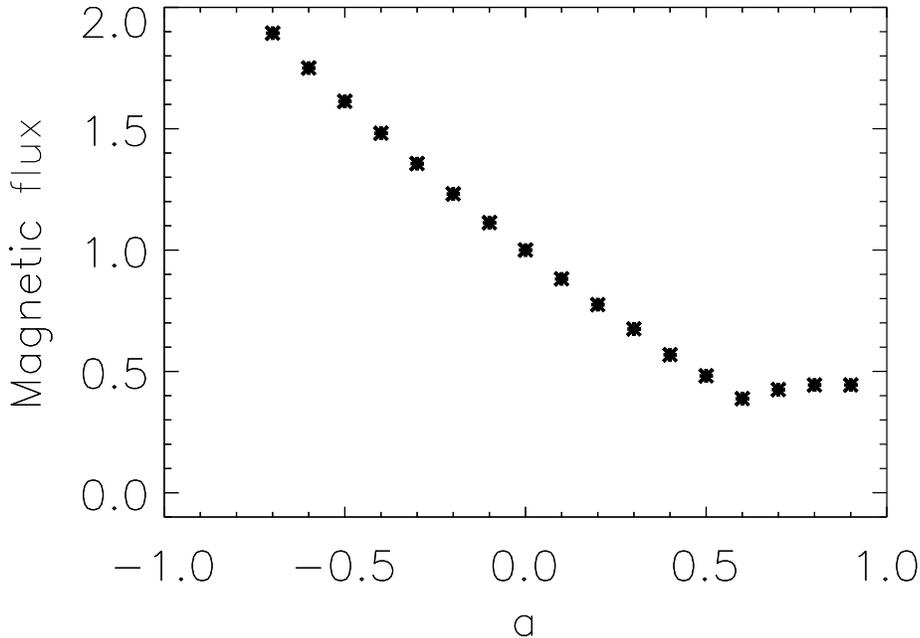}}
\caption{Magnetic flux through a polar cap on the horizon normalized
to the flux in the spin-zero case, vs. black hole spin illustrating
the basic result that flux-trapping is less effective as the spin
increases up to intermediate values of prograde spins.}\label{B_flux}
\end{figure*}

We evaluate the steady-state solution to the above equations for
various black hole spin values and for a range of magnetic Prandtl
numbers from $Pr_{m}=20$ to $Pr_{m}=2$, a disk thickness of $h/r=0.1$,
$r_{dead}=40r_{\rm g}$ and $r_{out}=53r_{\rm g}$. We find that the
flux accumulated on the black hole horizon decreases as the spin
increases from $-0.9$ up to about $0.6$, beyond which the flux is roughly
constant (figure \ref{B_flux}).

\begin{figure*}[h!] 
\centerline{\includegraphics[angle=-0,scale=0.8]{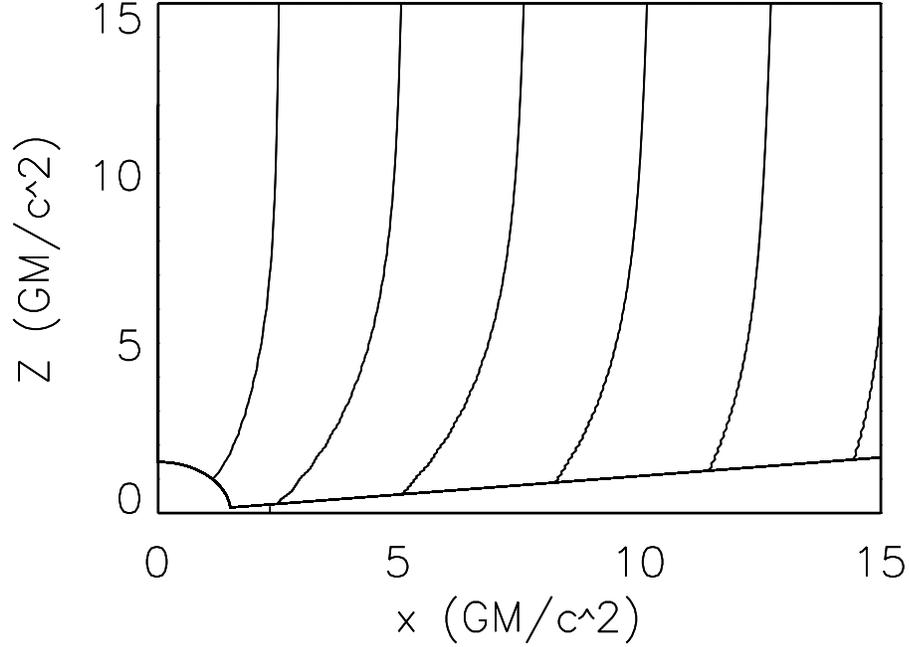}}
\caption{Magnetic flux configuration for black hole with $a=0.9$.
Each side is 15 Boyer-Lindquist radii.  The small vertical line at
$x\approx2.3$ indicates the marginally stable orbit, inwards of which
is the plunging region}\label{prograde}
\end{figure*}

\begin{figure*}[h!]
\centerline{\includegraphics[angle=-0,scale=0.8]{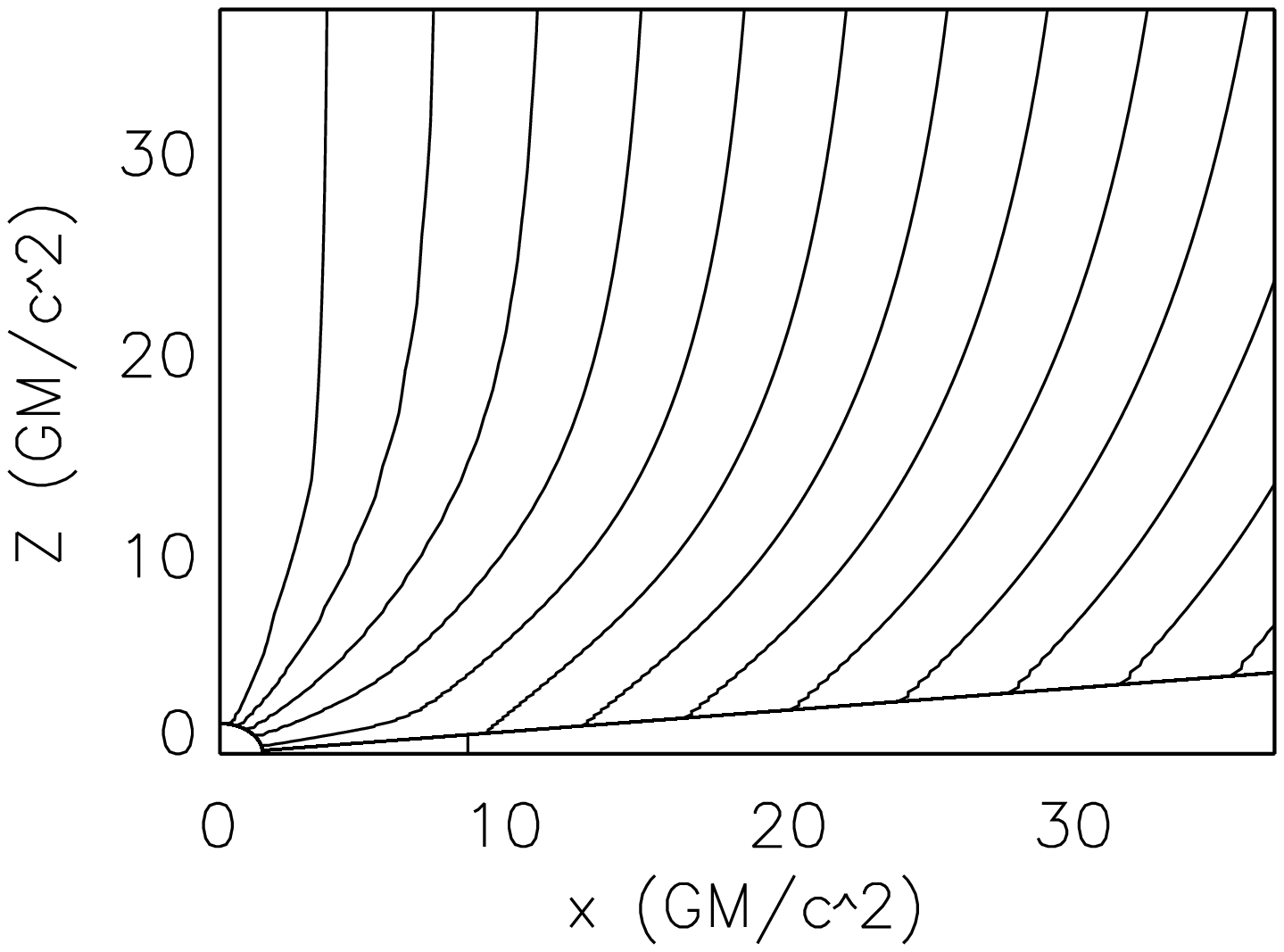}}
\caption{Lines of constant magnetic flux for $a=-0.9$ retrograde
spinning black hole.  The axes are in BL radii.  The marginally stable
orbit is located at about $x=8.7$.}\label{retrograde}
\end{figure*}

The reason behind the decrease in flux with increase in spin appears
to have a straightforward geometrical interpretation.  As the spin of
the hole increases, the accretion disks' inner edge (at the radius of
marginal stability) gets closer to the horizon in both coordinate and
proper distance.  This results in a decrease of the ratio of the area
within the plunge region to the area of the event horizon.  Thus, as
one considers more rapidly rotating black holes, the geometry becomes
progressively less favorable to shielding the turbulent accretion disk
from the hole-threading flux bundle.  We show the geometry of magnetic
flux lines for $a=0.9$ in Fig. \ref{prograde} and $a=-0.9$ in
Fig. \ref{retrograde}.

As mentioned above, the behavior of the radius of marginal stability
for rapidly prograde spinning black holes results in ineffective
shielding of the hole-threading flux bundle from the turbulent
(diffusive) portion of the accretion disk.  In short, flux-trapping
breaks down for largest prograde spin, so the magnetic field threading
the horizon is not enhanced and thus is equal in strength to that at
the disk inner edge.  Our runs, however, are done up to spin of 0.9
where the enhancement is smaller but still nonzero (i.e. the ability
of the plunge region to enhance the field has not dropped to zero).
Despite a decrease in size of the plunging region as spin increases,
the flux values do not drop further at spins above about $0.6$.
Therefore, the size of the plunging region does not solely determine
the dynamics of magnetic flux accumulation on the black hole; but, MHD
effects at higher prograde spins kick in to attempt to increase the
black hole-threading flux.

\begin{figure*}[h!]
\centerline{\includegraphics[angle=-0,scale=0.6]{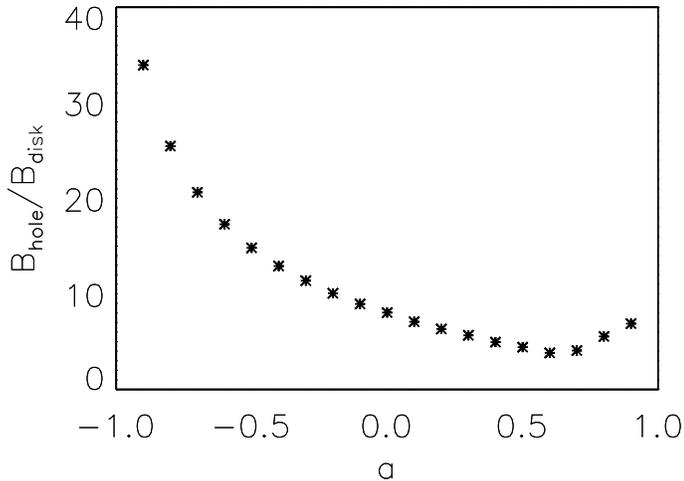}}
\caption{Ratio of horizon-threading magnetic field as measured by ZAMO
  observers and magnetic field strength in the accretion disk as a
  function of spin.}\label{B}
\end{figure*}

We now consider the BZ power that results from the trapped magnetic
flux and its dependence on black hole spin.  We start by evaluating the
horizon-threading magnetic field as measured by ZAMO observers from
the flux values we obtain,
\begin{equation}
\label{zamo_b}
B_{H}=\sqrt{g_{11}}B^{r}
\end{equation}
with 
\begin{equation}
B^{r} = \ast F^{rb}u_{b},
\end{equation}    
where $\ast F^{ab}$ is the dual Faraday tensor and $u^{b}$ is the
four-velocity of the ZAMO observers evaluated in the equatorial plane
on the horizon membrane (in the sense of Thorne et al, 1986).  The
dual tensor components involve terms with derivatives of $A_{\phi}$
with respect to $\theta$ and therefore require the use of the regularity
condition on the horizon membrane (eq. \ref{Znajek}).  The results
depend on the value of $B_{0}$, the initial uniform field strength
threading the horizon and are displayed in Figure \ref{B}.

\begin{figure*}[h!]
\centerline{\includegraphics[angle=-0,scale=0.8]{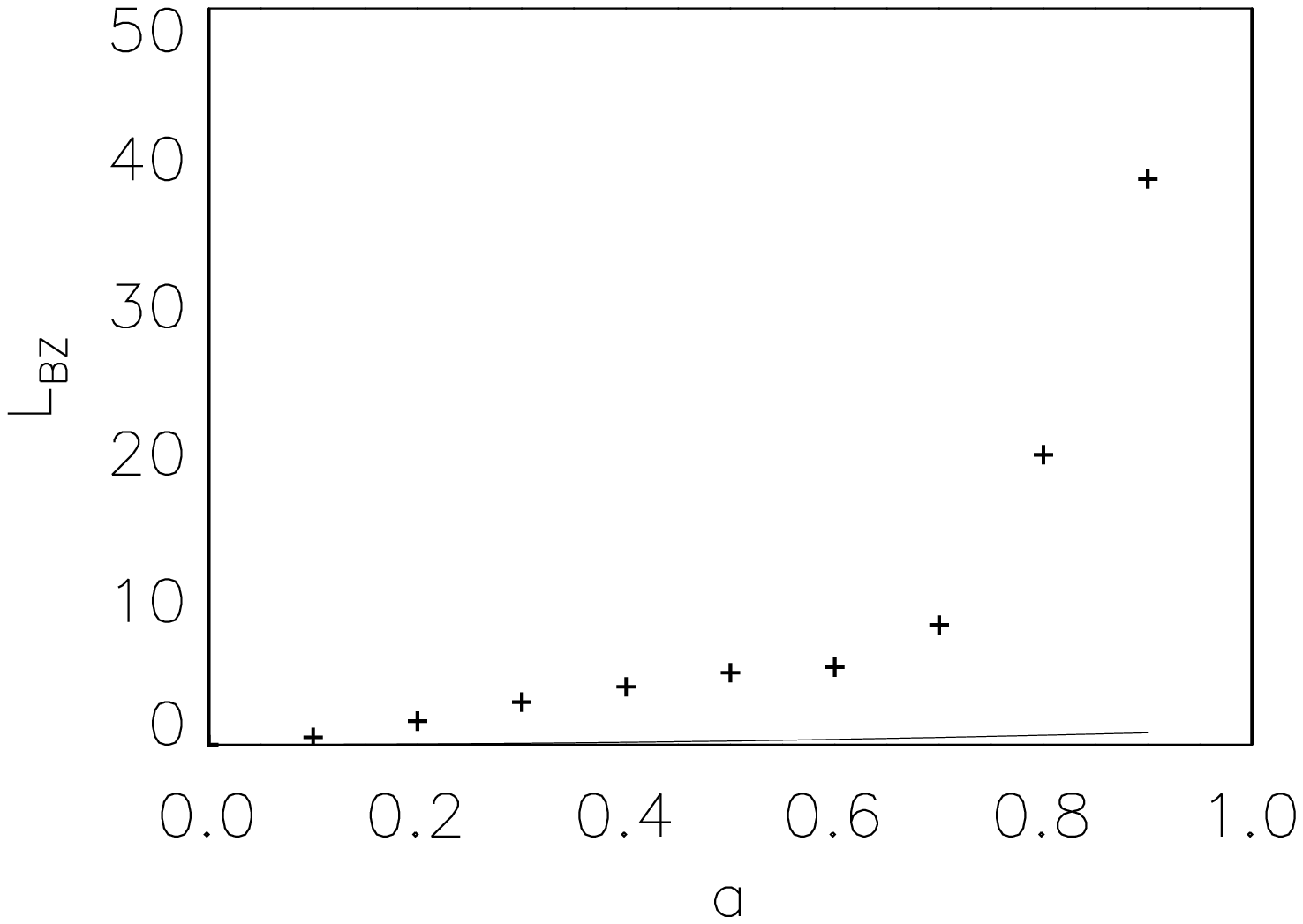}}
\caption{The individual points label the Blandford-Znajek luminosity
  as a function of spin for $Pr_{m}=20$.  The solid line is the
  ``classical'' BZ power.}\label{L}
\end{figure*}

With these values of the horizon-threading magnetic fields, we can
determine the BZ luminosity.  This is shown for prograde spins ($a>0$)
in Figure \ref{L}.

Despite possessing the largest hole-threading magnetic field for
non-negative spin, the Schwarzschild case, of course, produces no BZ
power.  Considering prograde black holes, the maximum BZ power is
generated for large $a$ as it is for the BZ mechanism without the
flux-trapping scenario.  Figure \ref{L} shows both the BZ power
without flux-trapping (straight line) as well as the BZ power obtained
in the flux-trapping model for a magnetic Prandtl number of 20.  We
note that the accumulated flux on the black hole depends linearly or
almost so, on the radial coordinate value of the outer region of the
active accretion disk.  We have truncated the active region of our
disk at $r_{dead}=40r_{g}$ outwards of which the disk fluid is
characterized by zero radial velocity so the dragging of flux occurs
only inwards of $r_{dead}$.  As pointed out in RGB, $r_{dead}$ is one
of the most artificial aspects of our model, but could be identified
with the outer edge of the MHD-turbulence dominated accretion disk, or
as the transition radius between an outer thin disk and an inner ADAF
disk.  We see, thus, that by increasing our active region of the disk
and/or increasing the disk thickness, we could increase the generated
luminosity.  However, we remind the reader that our analysis is
performed with large magnetic Prandtl number (i.e. 20) and thus with
low diffusion.  RGB shows that lower values of $Pr_{m}$ will generate
considerably lower black hole threading flux and this result is
confirmed in the relativistic regime as well.  The fundamental
difference between BZ power for high and low Prandtl numbers is a
shift in magnitude.  While the spin dependence is unchanged, the power
for $Pr_{m}=2$ is lower by about one order of magnitude.  One final
point is in order.  The outer boundary condition at $r=54r_{g}$ is
fixed which means that field dragging by the disk towards the black
hole will generate bending of field lines that is greater than an
analogous simulation in which the outer boundary is at a larger radial
coordinate.  Since the bending of field lines increases the diffusion
term in equation \ref{disk2}, the choice of fixed outer boundary not
only decreases the overall flux accumulation on the hole, it does so
more for smaller choice of outer radial coordinate value.  Like
$r_{dead}$, $r_{out}$ is an artificial aspect of our model whose
physical value might be interpretable as some kind of load region
where flux-freezing forbids flux lines from being dragged along with
the field threading the accretion disk.  In RGB, on the other hand, we
did not fix the field at $r_{out}$, thereby allowing the field to be
dragged unrestricted by the outer boundary value.  Nevertheless, these
caveats do not affect our basic qualitative result that the BZ
luminosity is greater in the flux-trapping model and has the spin
dependence shown.

\section{Discussion}
In the previous section we showed that the magnetic field threading
the black hole is always greater in the flux-trapping model which
leads to larger BZ powers.

We will now show that our model has immediate application to the
curious properties of jetted AGN in nearby elliptical galaxies.

Allen et al. (2006) used the {\it Chandra} X-ray observatory to study
nine nearby X-ray luminous elliptical galaxies.  Assuming central
black hole masses given by the M-$\sigma$ relation, {\it Chandra}
measurements of the ISM temperature and density of scales
$\approx10pc$ from the cores of these galaxies could be used to deduce
the rate at which ISM accretes into the gravitational potential of the
black hole.  These estimates are based on the simple spherical
accretion picture of Bondi (1952).  In addition, {\it Chandra} reveals
ISM cavities that have been blown by jet activity from the central
AGN.  Using ``PdV'' arguments (and assuming that the cavities have an
age given by their sound crossing time), the jet powers could be
deduced.  It was found that
\begin{equation}
\label{jetPower}
P_{jet}\approx\eta\dot{M}c^{2}
\end{equation}
where $\eta\approx 3$.  The object-to-object scatter about the
correlation (eqn. \ref{jetPower}) is small, with deviations in
efficiency of only a factor of $\approx 2$.  

Nemmen et al. (2006) have explored whether the Allen et
al. correlation is a natural result of the BZ mechanism.  Employing
the Narayan \& Yi (1995) advection dominated accretion flow (ADAF)
model, which is likely appropriate for the low accretion rates found
in these elliptical galaxies, Nemmen et al. estimated the strength of
the magnetic field in the central disk as a function of accretion rate
and then estimated the BZ efficiency, $\eta_{BZ}$, as a function of
black hole spin, where
\begin{equation}
\eta_{BZ}\equiv\frac{L_{BZ}}{\dot{M}c^{2}}.
\end{equation}
They showed that the Allen et al. results could only be reproduced if
the elliptical galaxy black holes were all rapidly spinning and there
was rather little mass loss in the accretion flow between the Bondi
radius and the black hole.  Although not as dramatic as in BZ, high
spins are required for the hybrid model as well (Meier, 1999).  Hence,
all of the black holes in the Allen et al. sample would need to
possess very close to the same, high prograde spin values; there is
clearly a fine tuning problem with these models.  For a review of the
current state-of-the-art in measurements of black hole spin and why
they should span a wide range, see Brenneman et al., 2009.
\begin{figure*}[h!]
\centerline{\includegraphics[angle=-0,scale=0.80]{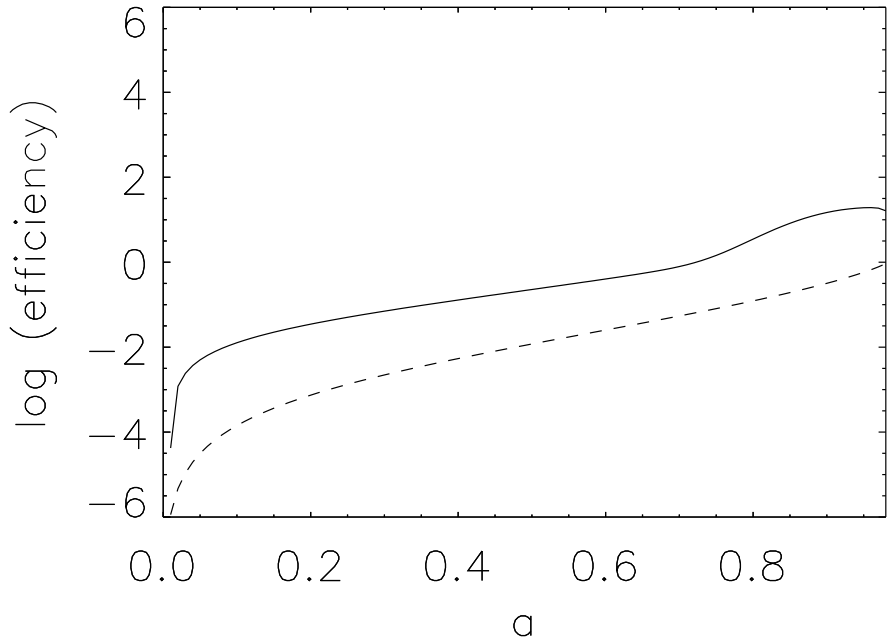}}
\caption[Nemmen]{The efficiency $\eta_{BZ}$ vs. spin for the BZ model
for a standard horizon-threading magnetic field (dashed line) vs. that
in the flux-trapping model (solid line).}\label{Nemmen}
\end{figure*}  

Assuming that the black holes in these elliptical galaxies are
rotating in the prograde sense, the flux trapping model alleviates
this fine tuning problem.  We have taken the ADAF model of Nemmen and
included the effects of flux trapping.  The resulting $\eta_{BZ}$ as a
function of $a$ for $a>0$ is shown in Fig. \ref{Nemmen} and compared
with that of Nemmen et al.  We caution the reader to ignore the solid
line beyond $a=0.9$ as it is simply a continuation of the fit that
captures the behavior for lower spins.  The absolute power that can be
produced by the BZ mechanism in the flux trapping model is
substantially higher (and slightly higher than that in the hybrid
model); thus we can tolerate a much larger mass loss between the Bondi
radius and the black hole.  In addition, for intermediate values of
prograde spin, the efficiency of the solid line representing the
flux-trapping model is a flatter function of spin than the dashed
line, producing a range of spin values compatible with the Allen et
al. results.  This $\eta_{BZ}$ results from $Pr_{m}=20$.  For lower
values of the Prandtl number down to $Pr_{m}=2$, although the spin
dependence of $\eta_{BZ}$ is unchanged, the curve shifts downward.
This forces the spin values of the black holes in the 9 AGN to be
above about 0.6 as opposed to being above about 0.4 for $Pr_{m}=20$.
However, we remind the reader that in addition to the Prandtl number,
the disk thickness and size also factor into the flux accumulation on
the black hole.  In fact, we could use $Pr_{m}=2$ and increase the
disk dead zone to $400$ as opposed to $40$ and get about the same
magnitude of flux on the black hole.  Alternatively, we could use
$Pr_{m}=2$, decrease the outer boundary to $200$ and increase the disk
thickness to $0.2$ and again produce similar magnetic flux on the
black hole.  In fact, it would not be unreasonable to achieve greater
magnetic flux values on the hole using $Pr_{m}=2$ instead of
$Pr_{m}=20$ but increasing the other parameters in the
disk.  Ultimately, we suggest that this provides a more compelling
exploration of the Allen et al. correlation than the more standard BZ
model of Nemmen et al.

\begin{figure*}[h!]
\centerline{\includegraphics[angle=-0,scale=0.8]{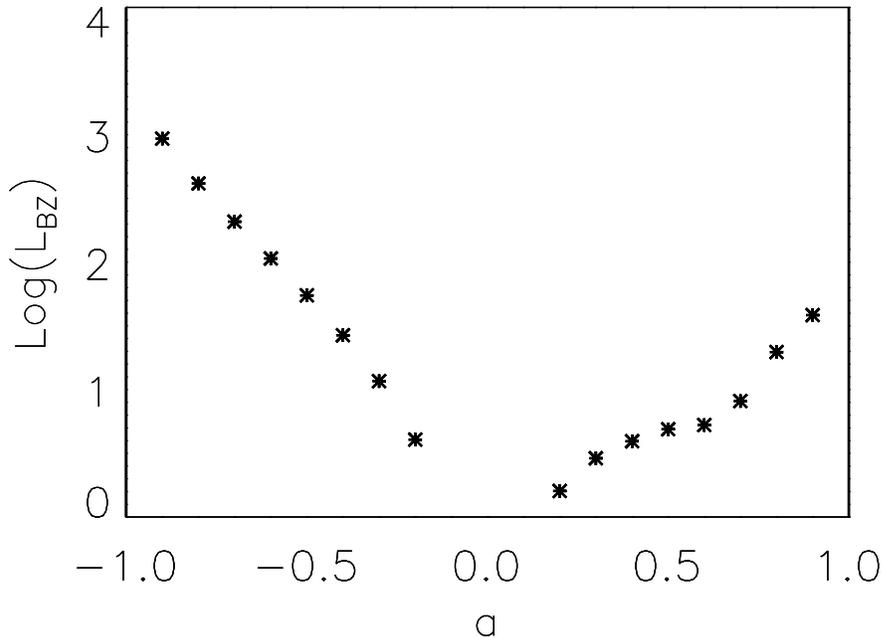}}
\caption{Blandford-Znajek luminosity with $Pr_{m}=20$ for the entire
range of spins showing that for retrograde values the power can be
more than one order of magnitude greater.}
\label{L2}
\end{figure*}


Finally, we note another interesting and novel prediction of the
flux-trapping model.  All other things being equal, accretion onto a
rapidly rotating {\it retrograde} black hole will result in
substantially higher BZ luminosities than possible in the prograde
case (see Fig.~\ref{L2}).  This is a direct consequence
of the fact that the radius of marginal stability around a black hole
that is rotating in a retrograde sense compared with the accretion
disk will be {\it larger} than the corresponding Schwarzschild value,
thereby leading to an even more dramatic ``geometrical'' enhancement
of the magnetic flux trapped on the black hole.  We suggest that some
of the most powerful radio galaxies possess jets that are energized by
accretion onto retrograde black holes.  We note that recent {\it
Suzaku} observations of the powerful broad-line radio galaxy 3C~120
found a broad iron emission line with a profile indicating an inner
disk truncation at $r=8.6^{+1.0}_{-0.6}r_g$ (Katooka et al. 2007),
consistent with the location of the radius of marginal stability for a
maximally rotating retrograde black hole ($r_{\rm ms}=9r_g$).  

\section{Comparison to recent GRMHD simulations}
In this section we point out and comment on the fact that our results
appear incompatible with recent GRMHD simulations.  In McKinney \&
Gammie (2004), for example, the authors find that the powers for
retrograde spins do not exceed the prograde ones and in GRMHD
simulations, in general, the magnetic field on the black hole
increases with larger prograde spin (e.g. De Villiers, 2005; Hawley \&
Krolik, 2006; McKinney, 2005) so there is no initial drop with lower
prograde spins.  The reasons for this difference appear to be twofold.
GRMHD applies the flux-freezing condition (ideal GRMHD), which means
that the magnetorotational instability (MRI) acts as a dynamo closer
in to the black hole as the spin is more prograde since the inner edge
of the disk is closer to the black hole.  The validity of this result
goes to the heart of the issue of physical vs. numerical resistivity
and whether or not ideal GRMHD is less diffusive than realistic black
hole flows.  Within the context of ideal GRMHD, it is not surprising
that retrograde BZ powers are not larger than prograde powers.  The
MRI operates further out for retrograde disks compared to the prograde
ones and thus is incapable of increasing field strengths in the inner
region.  We should highlight a feature of retrograde powers in our
model that may seem puzzling.  In order to avoid large discontinuities
in the field line angular velocity in the transition region between
the disk and the hole, we are forced to an angular velocity
prescription that lowers BZ powers for all values of spin but in a way
that increasingly lowers the power as the spin becomes more
retrograde.  This is because the discontinuity that would arise
between the horizon and inner accretion disk is greater as the spin
becomes more retrograde, so that avoiding a discontinuity has a
greater effect on retrograde powers.  But the plunge region boundary
condition comes to the rescue, so to speak, in that it counteracts the
otherwise large decrease in power that would result in large
retrograde systems that stems from this smoothening out of the angular
velocity profile in the transition region.  The instantaneous
advection assumption for the plunge region, in fact, produces
sufficiently strong field on the black hole to compensate for the
otherwise decrease in power.  If we ignored the discontinuities and
enforced an angular rotation of field lines on the horizon that is
rigidly that of the black hole and one in the inner disk that is
rigidly Keplerian, the powers for increasing retrograde spins would be
even larger than shown.  That solution, however, would not be
self-consistent.  In short, the plunge region boundary condition is
fundamental in appreciating the difference between the flux-trapping
model and ideal GRMHD.  We should also point out that whereas our
model is by assumption force-free everywhere outside the disk and
plunge region plane, this condition exists only in specific regions in
ideal GRMHD simulations.  Specifically, the lack of a force-free
configuration everywhere above and below the plunge region plane in
GRMHD, means that the inertia of the gas is not negligible and
therefore weighs in on the issue of magnetic flux accumulation on the
black hole.  Although the ability of the plunge region to enhance the
field on the hole is weaker in GRMHD, it is nonetheless confirmed (for
example, McKinney \& Gammie, 2004).  In RGB, it was shown that the
enhancement of the hole-threading field with respect to the magnetic
field in the inner disk depends not only on physical quantities such
as disk Prandtl number but also on geometric aspects of the magnetized
disk such as length and thickness.  In addition, the enhancement on
the hole compared to the inner disk depends also on the geometry of
the initial magnetic field configuration ( De Villiers et al, 2005;
McKinney \& Gammie, 2004).  In other words, if the numerical solution
in ideal GRMHD is less diffusive than the physical solution, and the
initial conditions are such that the plunge region enhancement is
small, then the magnetic field on the black hole can increase as the
spin increases for all prograde values.  In short, the issue of ideal
vs. non-ideal GRMHD, as well as the issue of the initial conditions
for the disk parameters and magnetic field, may conspire in ideal
GRMHD to hide not only the effectiveness of the plunge region to
enhance the B-field on the black hole compared to the inner accretion
flow, but the spin dependence of this process as well.  These issues
cannot be addressed within the confines of our simple model and we
limit our discussion, thus, to emphasizing that it may be more
appropriate to compare our study to a future one that includes large
scale magnetic fields treated within resistive GRMHD.  In addition,
one should keep in mind that this study is carried out in the thin
disk approximation which is only assumed to occur for intermediate
accretion rates.  Furthermore, even for those intermediate accretion
rates, it has recently been shown that thin-disk accretion does not
necessarily have to result (Fragile \& Meier, 2009).  Until radiative
processes as well as physical resistivity are included, a realistic
assessment of the detailed interaction between black holes and
magnetized flows cannot be made.  The effects discussed in this paper
involve the spin dependence in a regime in which the dynamics of the
plunge region is dominant, and in which the magnetized accretion flow
does not act as a dynamo.

\section{Conclusions}
\label{conclusion}
We postulated that given the dynamics of the plunge region of a thin
black hole accretion disk, flux trapping can enhance the strength of
the magnetic field threading the horizon by a significant factor.  If
the trapping behavior of the plunge region operates in as dominant a
fashion as this study assumes, and if the energy emitted is BZ
luminosity, one finds that the efficiency is a fairly flat function of
spin for intermediate values of black hole spin.  This allowes one to
follow the program of Nemmen et al to show that the enhancement due to
the flux-trapping model on the BZ power is sufficient to explain the
energies of the nine jets in the Allen et al sample including an
attractive indeterminacy in the black hole spin.  Finally, we suggest
that some of the most powerful radio galaxies possess jets that are
energized by accretion onto retrograde black holes.

\section*{Acknowledgments}

The author thanks Christopher S. Reynolds as my graduate advisor for
providing the opportunity to work on an interesting topic, for
extensive discussion and for suggesting the application of the
flux-trapping model to the Allen et al data.  I also thank Ted
Jacobson for helpful discussion on the relativistic equations, Cole
Miller for useful suggestions on the draft as a whole, and David
L. Meier for highlighting features of retrograde black hole accretion.
Finally, I thank the anonymous referee for fundamental changes.
D.G. is supported by the NASA Postdoctoral Program at NASA JPL
administered by Oak Ridge Associated Universities through contract
with NASA but also acknowledges partial support from NSF grant
AST0205990.


\section*{References}

\noindent  Allen, S.W., et al., 2006, MNRAS, 372, 21

\noindent  Balbus, S. A., \& Hawley, J.F., 1991, ApJ, 376, 214

\noindent  Blandford, R. D., Payne, D.G., 1982, MNRAS, 199, 883     

\noindent  Blandford, R. D., \& Znajek, R. L. 1977, MNRAS, 179, 433 

\noindent  Bondi, H., 1952, MNRAS, 112, 195

\noindent  Brenneman, L. Astro2010 Science White Paper

\noindent  Chandrasekhar, 1961, Hydrodynamic and Hydromagnetic
           Stability, Oxford University Press.

\noindent De Villiers, J.P., Hawley, J.F., \& Krolik, J.H., 2003, ApJ,
599, 1238

\noindent De Villiers, J.P., Hawley, J.F., Krolik, J.H., Hirose, S.,
2005, ApJ, 620, 878

\noindent Fragile, P.C., Meier, D.L. 2009, ApJ, 693, 771 

\noindent  Ghosh, P., Abramowicz M.A., 1997, MNRAS, 292, 887

\noindent  Hirose, S., Krolik, J.H., De Villiers, J.P. \& Hawley, J.F., 2004, ApJ, 606, 1083

\noindent Komissarov, S.S., 2001, MNRAS, 326, 41

\noindent  Livio M., Ogilvie G.I., Pringle J.E., 1999, ApJ, 512, 100

\noindent  Lubow, S.H., et al., 1994, MNRAS, 268, 1010


\noindent  Macdonald, D.A., 1984, MNRAS, 211, 313

\noindent  McKernan, B., et al., 2007, MNRAS, 379, 1359

\noindent  McKinney, J.C., Gammie C.F., 2004, ApJ, 611, 977

\noindent  McKinney, J.C., 2006, MNRAS, 368, 1561

\noindent  Meier, D.L., 1999, ApJ, 522, 753

\noindent  Meier, D.L., 2001, ApJ, 548, L9

\noindent  Moderski R., Sikora M., Lasota J.-P., 1998, MNRAS, 301, 142

\noindent  Narayan, R., Yi, I., ApJ, 1995, 452, 710

\noindent  Nemmen R.S., et al., 2007, ApJ, 377, 1652

\noindent  Reynolds, C.S., et al, 1996, MNRAS, 283, 873

\noindent  Reynolds, C.S., \& Begelman, M.C., 1997, ApJ, 487, 135 

\noindent  Reynolds, C. S., Garofalo, D., \& Begelman, M. 2006, ApJ, 651, 1023

\noindent  Thorne, K. S., Price, R. H., \& Macdonald, D. A. 1986, Black Holes: The Membrane Paradigm (New Haven: Yale Univ. Press)

\noindent  Wardle, J.F.C, et al, 1998, Nature, 395, 457

\noindent Znajek, R.L., 1978, MNRAS, 185, 833


\noindent Velikhov, F., 1959, Soviet Phys-JETP, 36, 1398

\end{document}